\documentclass[twocolumn,fleqn,10pt]{article}
\usepackage{graphicx}
\usepackage{color}

\usepackage[square,numbers,sort&compress]{natbib}
\usepackage{amsmath,amssymb,amsfonts}

\usepackage{graphicx}
\usepackage{array}
\usepackage{threeparttable}
\usepackage{booktabs}
\usepackage{tabularx}
\usepackage{color}
\usepackage{longtable}

\usepackage{titlesec}
 \usepackage{url}
\usepackage{times}
\usepackage{mathptmx}

\usepackage[font={small}]{caption}
\usepackage{amsmath,bm}

\title{Quantification of the underlying mechanisms and relationship among cancer, metastasis and differentiation/development}

\author{Chong Yu,$^{1}$ Qiong Liu,$^{1}$ Cong Chen,$^{3}$ and Jin Wang$^{1,2,3*}$\\
\normalsize{$^1$ State Key Laboratory of Electroanalytical Chemistry } \\
\normalsize{Changchun Institute of Applied Chemistry, Chinese Academy of Sciences } \\
\normalsize{Changchun, Jilin 130022, China } \\
\normalsize{$^2$ College of Physics} \\
\normalsize{Jilin University, Changchun, Jilin 130012, China}\\
\normalsize{$^3$ Department of Chemistry, Physics \& Applied Mathematics} \\
\normalsize{State University of New York at Stony Brook } \\
\normalsize{Stony Brook, NY 11794-3400, USA } \\
\normalsize{$^{*}$Corresponding Authors: jin.wang.1@stonybrook.edu}}
\date{}

\begin{document}
\maketitle



\newpage


\begin{abstract}
Recurrence and metastasis have been regarded as two of the greatest obstacles for curing cancer. Cancer stem cell (CSC) have been found. They contribute to cancer development with the distinct feature of recurrence and resistance to the popular treatments such as drugs and chemotherapy.  In addition, recent discoveries suggest that the epithelial mesenchymal transition (EMT) is an essential process in normal embryogenesis and tissue repair, which is a required step in cancer metastasis. Although there are many indications showing the connections between metastasis and stem cell, researches often studied them separately or at most bi-laterally, not in an integrated way. In this study, we aim at exploring the global mechanisms and interrelationship among cancer, development and metastasis which are currently poorly understood. To start, we constructed a core gene regulatory network motif which contain specific genes and microRNAs of CSC, EMT and cancer. We uncovered seven distinct states emerged from the underlying landscape. They are identified as Normal, Premalignant, Cancer, stem cell (SC), cancer stem cell (CSC), Lesion and Hyperlasia state. Given the biological definition of each state, we also discussed the metastasis ability of each state. We show how and which types of cells can be transformed to a cancer state and the connections among cancer, CSC and EMT. The barrier height and flux of the kinetic paths are explored to quantify how and which cells switch stochastically between the states. Our landscape model provides a quantitative way which reveals the global mechanisms of cancer, development and metastasis.
\end{abstract}

\section{Quick guide to equations and assumptions }
In a bio-chemical system, reactions will happen anytime. So the variations of the system states must be a complex process of dynamics. There are $M$ reactions ${R_1,R_2,...,R_M}$ of the protein regulations. Each reaction $R_j$ corresponds to a propensity function $a_j$. $a_j(x)dt$ is defined as: to a given $X(t)=x$, the probability of $R_j$ has reacted once within time $[t,t+dt)$. In general, it is hard to describe the propensity function $a_j$ accurately. We can obtain an approximate description: $a_j(x)= c_j\prod_{k=1}^{N}C^{m_{j_k}}_{x_k}$.

$\prod_{k=1}^{N}C^{m_{j_k}}_{x_k}$ is the probability of the molecular $S_k$ will participate in the reaction $R_j$. The value of $x_{k}$ represents total number of $S_k$, and ${m_k}$ represents the number of reactants which participate in the reaction. $c_j$ is the constant of chemical reaction, which can obtain from experiments.

In our model, we use Gillespie algorithm\cite{gillespie1977exact} to obtain the stochastic distribution time of protein binding/unbinding. Due to intrinsic fluctuations and variations of the protein molecule numbers in the cells, we explore the stochastic dynamics time of each protein binding/unbinding steps and change the protein molecule numbers correspondingly as $c_j\prod_{k=1}^{N}C^{m_{j_k}}_{x_k}$.

In order to describe the biological process precisely, we defined a set of time scale parameters of each process. We used the parameters $g$ for protein synthesis rate and $k$ for protein degradation rate, $h$ is the binding rate and $f$ is the unbinding rate of regulatory proteins to the target genes. The protein synthesis rate is influenced by the regulated gene number and regulated type. There are two regulated type: binding state and unbinding state. If the protein has $n$ binding site, the protein will has $2^n$ synthesis rates. The synthesis rate will be increase by a factor of $\lambda_a$ or decreased by a factor of $\lambda_r$, respectively. If there are 2 binding sites one is activation the other is repression, the 4 synthesis rates will be set as: $g_{00}$, $g_{01}$ =  $g_{00}\lambda_a$, $g_{10} =  g_{00}\lambda_r$, $g_{11} =  g_{00}\lambda_a\lambda_r$. We define the equilibrium constants: $X_{eq} = f/h$, the adiabatic parameter: $\omega = f/k$. The later is used to quantify the unbinding time of a protein in its lifetime. If the value of $\omega$ is large, it means the regulation processes are relatively fast compared to the synthesis/degradation which is sometimes termed as adiabatic. If the value of $\omega$ is small, it means the regulation processes are relatively slow. The protein switches on and off to the target gene relatively slowly which is non-adiabatic. Please see the SupportInformation for details. In this work, we mainly discussed the adiabatic case.
\section{Introduction}
Cell phenotypes change during the development of cellular differentiations \cite{Wang2011-8257,Xu2014Exploring}. Differentiation starts from an oosperm which develops into a complex system of biont and continues in adulthood as stem cells divide and generate differentiated daughter cells when the tissue repair and cells regenerate\cite{Sell2004Stem}. Induced pluripotent stem cells (iPS) are a type of pluripotent stem cells which provide the
opportunity of the therapeutic uses\cite{Takahashi2007-861}. The adult cells have been reprogramed into pluripotent stem cells in 2006 \cite{Takahashi2006-663}. This is a significant step in the stem cell and regenerative biology as the cell type switching can skip many intermediate steps. This lineage reprogramming technology may also have profound implications for cancer biology.

Cancer has been one of the most deadly disease for human beings. Studies show that there are multiple factors associated with the recurrence and metastasis which lead to cancer fatal to human beings\cite{Cowin2005Cadherins}. Many researches concentrate on the origin of cancer being from the genetics (mutations)\cite{Muller2013p53,Martincorena2015Somatic}. The accumulation of mutations led to malignant transformation which was described as a disease of clonal evolution. Through these mutation and selection, the cells acquire the hallmarks of cancer\cite{Hanahan2000-57,Lynch1998Mutation}. Some cells may acquire hypoxia characteristic. Some cells may acquire fast-growing characteristic. Some cells may develop new blood vessels and so on. This is a widely accepted concept for understanding the generation of cancer. On the other hand, more and more observations have demonstrated that cancer should be thought of an intrinsic state which emerges from the underlying gene regulation networks\cite{Kauffman1971Differentiation,Spano2012Molecular}. The gene regulation network controlled a series of cellular activity and biological processes. The network can perform the regulation instructions which may affect early events of cancer\cite{Blancafort2013Writing}. These network environmental and epigenetic effects can result not only silencing of tumor suppressors but also reactivating the silenced regions which could prime subsequent events in the development of cancer\cite{Rodr2011Cancer,Liu2008BRCA1}.

Cancer stem cells (CSCs) can be defined as the cells with the characteristics of cancerous and stem cell-like features\cite{Tang2012Understanding}. The CSCs are considered to be the seeds of cancer. Although cancer cells might be killed during the radiation and chemotherapy, the CSCs as seeds of cancer can still survive. This can explain the cancer recurrence after treatments. Although the CSC theory has been reported as early as 1952\cite{hewitt1952transplantation}, the importance of that had been realized recently. CSCs have been found that they serve as the basis of cancer developement, maintenance, metastasis and recurrence\cite{Dragu2015Therapies}. In general, the development/differentiation process is from the primary stem cell and the reprogramming is vice versa which is important to the tissue reengineering. The example of cellular reprogramming is induced pluripotent stem (or iPS) cells, which gives the hope for cell fate switching and transformation\cite{Kondo2000Oligodendrocyte}. However reprogramming often encounters cancer state which result to the transformed progenitors acquiring self-renewal and cancerous characteristics\cite{Pavlova2016The}. One may encounter the CSCs. Furthermore, CSCs facilitate the primary tumor cells to migrate from one location to another which is a key step in the metastatic cascade.

EMT is an essential process, through which most adult tissues maintain the migratory capacity in normal embryogenesis, wound healing and tissue repair\cite{morel2012emt}. CSCs can plant into another organ also through the EMT process\cite{wang2015links}. In the EMT process, a set of transcription factors(TFs) induce the early steps of metastasis\cite{scheel2012cancer}. Through the EMT-TFs, the differentiated epithelial cells can obtain mesenchymal traits to colonize foreign tissues and create new tumor site in distant organs. Moreover, EMT process is also a trail for non-stem cells turning into  stem  cell  states. Experiments have observed that inducing an EMT process during normal mammary epithelial cells differentiation can make a generation of mammary epithelial stem-like cells\cite{mani2008epithelial}. This kind of experimental phenomena can be observed in both normal and cancerous tissues\cite{morel2008generation}. Thereby, EMT is an important process which not only contribute to creating metastatic CSCs, but also has a close relationship with CSCs\cite{chaffer2011normal}.

Despite many evidences showing the connections between the metastasis and stemness of the cell, or cancer and differentiation/development \cite{li2015quantifying}, these are still rarely studied in an integrated way of cancer, metastasis and development/differentiation. We aim to explore the connections among cancer, development/differentiation and metastasis in a systematic and quantitative way. We start by constructing a core gene regulation network motif. In order to characterize the key points of the dynamic process, the specific genes and microRNAs of cancer, CSC and EMT was included. In this work, we quantified the underlying landscape of cancer, metastasis and development/differentiation. Furthermore, we include regulatory binding/unbinding information to make the model more precise. Seven states emerged from the landscape which are quantified by the basins of attractions representing the Normal, Premalignant, Cancer, SC, CSC, Lesion and Hyperplasia states. We define these states by the gene expression level and the biological significance. We discuss the metastasis ability of these states as well. There are three kinetic paths from Normal to Cancer state. Two kinetic paths which connect CSC state show the formation of cancer stem cells from two sources. The optimal paths and barrier height between the states can illustrate how and which cells will be able to transform to cancer state and why cancer is so difficult to cure. This leads to a quantitative understanding of the degree of difficulty in curing the cancer. Moreover, the quantified landscape give us a portrait which uncover the dynamic interrelationship of the biological process among CSCs, EMT and cancer. Then we use global sensitivity analysis to discuss which regulation is more sensitive for cancer curing which can give a guide on clinical experiments of cancer. This work can elucidate the origin of cancer, as well as the process of cancer development/differentiation and metastasis. This has clear clinical significance in helping to understand the CSC basis of treatment response, therapeutic resistance, and cancer relapse.

\section{Results and Discussion}
\subsection{Model Construction}

\begin{figure}[!ht]
\includegraphics[width=0.50\textwidth]{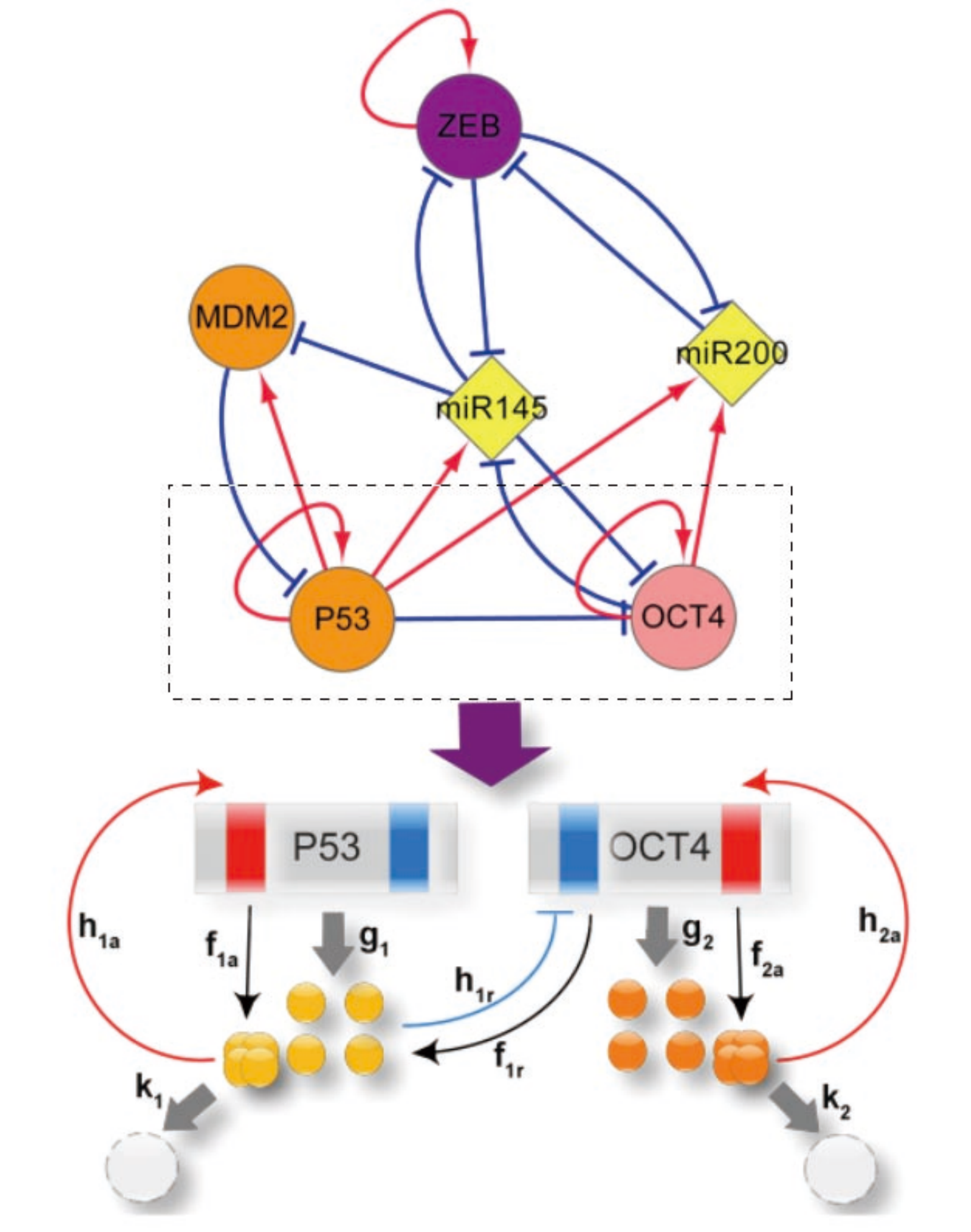}
\caption{The diagram for the core gene regulatory network motif which contain 6 nodes and 16 regulations (7 activations and 9 repressions, the arrows represent the activate regulations, the short bars represent the repress regulations). The nodes of diamond shape represent the microRNAs. The orange round ones represent the specific genes of cancer, the violet one represents the specific gene of EMT, and the pink one represents the specific gene of CSC.} \label{net}
\end{figure}

To emphasize the characteristics of CSCs, EMT and cancer, a core gene regulatory network motif covers specific genes and micro-RNAs of the three aspects. As shown in Fig.\ref{net}, MDM2 is an oncogene of cancer, P53 is a well known tumor suppressor gene\cite{yu2016physical}. ZEB is an EMT activator gene which suppresses the stemness-inhibition of micro-RNA (mir-200)\cite{wellner2009emt}. OCT4 is an essential gene which mediates phenotype self-renewal and stemness\cite{kumar2012acquired}. mir-145 and mir-200 are two important microRNA which play vital roles in both CSC and EMT regulation\cite{liu2015epigenetic}. The arrows represent activation and the short bars represent the repression.

\subsection{Definition and metastasis ability of each steady states and the kinetic paths of the landscape}
There are 6 nodes in our network motif. It is difficult for us to visualize a 6-dimensional space. Thus we chose to discuss three specific genes P53, ZEB and OCT4, reflecting the cancer, EMT and development/differentiation (with CSCs) aspects. P53 is a tumor suppressor gene. The cells in normal function often with high gene expression level of P53. Low gene expression level of P53 is a general characteristics of cancer\cite{Yang2013-impact, yu2016physical}. OCT4 is a signature gene of stem cell. Many studies show that OCT4 is critically involved in the self-renewal and is a critical gene for cell differentiation and reprogramming \cite{lin2012reciprocal}. High gene expression level of OCT4 means the cells are of self-renewing ability, multi-differentiating potential, and strong proliferative ability. ZEB is a critical gene of EMT process. The expression of ZEB can activate EMT process and EMT is a required step in metastasis\cite{lamouille2014molecular}. The gene expression level of ZEB is a metastasis signature.

\begin{figure}[!ht]
\includegraphics[width=0.50\textwidth]{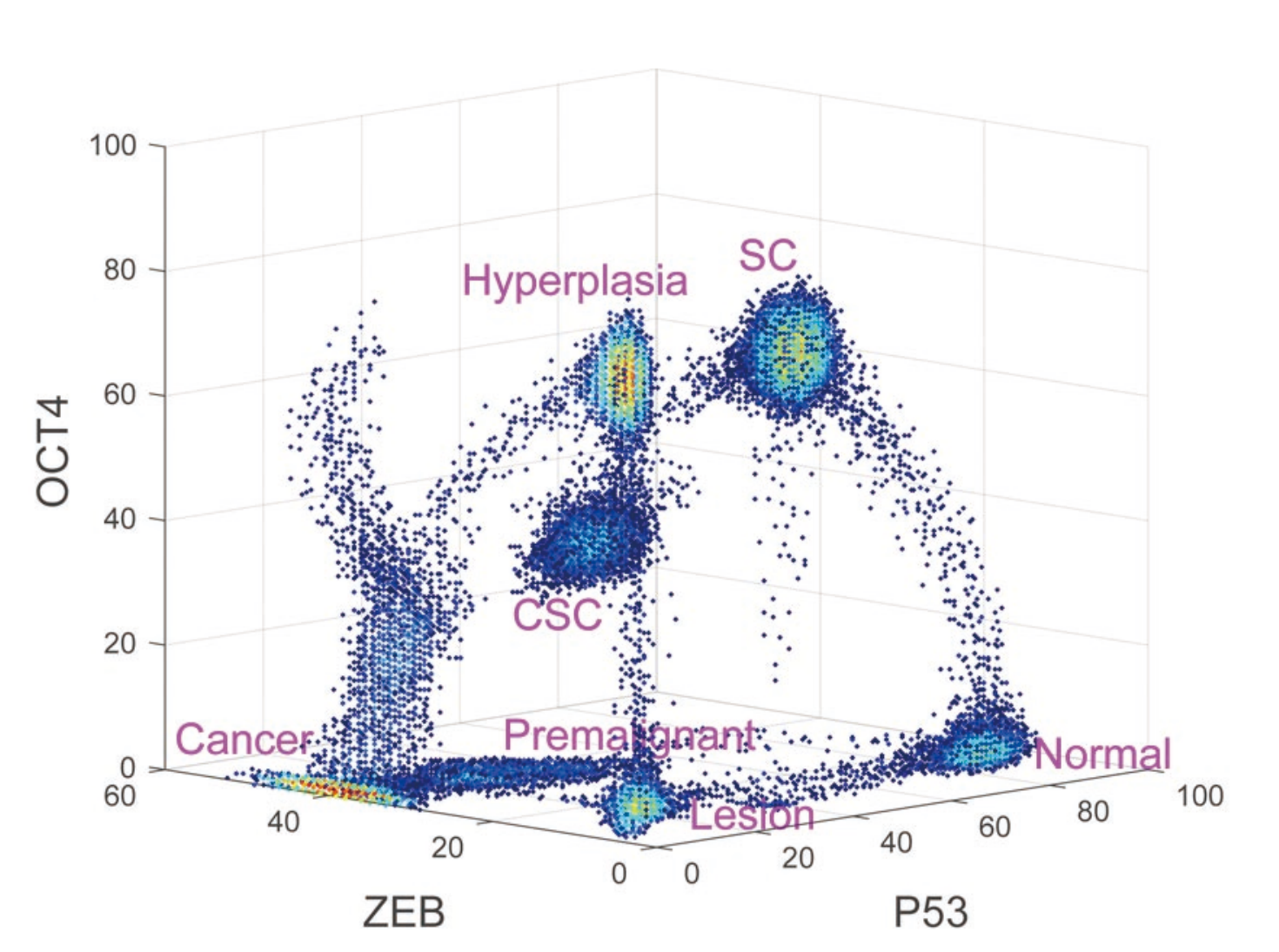}
\caption{The 3-dimensional landscape shows the  Normal, Premalignant, Cancer, SC, CSC, Lesion and Hyperplasia states and optimal paths among those states.} \label{3d}
\end{figure}

From Fig.\ref{3d}, we can see that there are seven states emerging which are named as Normal, Premalignant, Cancer, SC, CSC, Lesion and Hyperplasia, respectively. In Normal state, the gene expression of P53 is high and the gene expression of OCT4 and ZEB is low. This illustrates that if the cells stay in normal state, they have a normal function. They do not have the characteristics of stem-cells such as self-renewal or reprogramming, and do not have the metastasis ability either.

The Lesion state has a low expression levels of P53, OCT4 and ZEB. Low gene expression level of OCT4 and ZEB indicate that the cells do not have the characteristics of stem-cell or metastasis ability. Low gene expression level of P53 illustrate the cells do not have normal function which maybe caused by inflammation, pH, hypoxic and so on\cite{Jeremy2003Inflammatory}.

In Hyperplasia state (the tumor state without metastasis), compared with the Lesion state, the expression level of OCT4 is high and the expression level of P53 and ZEB are low. High gene expression level of OCT4 implies that the cells have the characteristics of stem-cells such as self-renewal or reprogramming. Hyperplasia state can be seen as the accumulated cell damage while the tissue which are inflamed starts the self-repairing which helps to produce new cells to replace the pathological cells. In this process, OCT4 is also a significant player in self-repair and DNA replication\cite{rizzino2013concise}. Low gene expression level of ZEB indicates that the metastasis is not significant. Low expression level of P53 indicates that the cells are still in abnormal condition. Cells in the Lesion and Hyperplasia states are both with a degree damage as the gene expression level of P53 is low. In general, they can be reversed to normal state by our self-healing system as the expression level of ZEB is low, the metastasis has not start yet.


In cancer state, the gene expression level of ZEB is high, the gene expression level of OCT4 and P53 is low. As we known, for the cells in cancer state, the P53 which is a tumor suppressor gene is in low expression level but the metastasis ability is obvious (high gene expression level of ZEB). Moreover, if the cancer cells are in the terminally differentiated stage, they have lost the ability to proliferate or to alter its destiny, the stemness ability is relatively low as well. So the gene expression level of OCT4 is low.

The Premalignant state (tumor state with certain metastasis) is a transition state between Normal and Cancer state. In the Premalignant state, the expression level of P53 becomes lower and the gene expression level of ZEB becomes higher when the cells transform from Normal to Cancer state. That means, when the cell state moves from Normal to Cancer, the cancerization and metastasis become more and more significant. As the metastasis ability of Premalignant state is intermediate. It has certain metastasis capability which brides the normal state and the complete cancer/metastasis state.

In the SC state, the gene expression level of P53 and OCT4 is high, ZEB is low. The SC state has the stemness activity so the expression level of OCT4 is high. The expression level of P53 is high and the expression of ZEB is low which means the metastasis ability is not active.

In the CSC state (the tumor state with significant stemmed and metastasis), the expression level of P53, OCT4 and ZEB are all in the intermediate level. The cells are in the transition between SC and Cancer state. The CSCs show up some characteristics of cancerization and self-renewal (stemness) as the gene expression level of P53 is lower and OCT4 is higher than the Normal state. Moreover, the elevated gene expression of ZEB shows that the cells have a certain ability of metastasis between the SC and Cancer state.

For the landscape perspectives, there are several major kinetic paths we can quantitatively explore. When the expression level of ZEB increases, the paths from SC to CSC and CSC to Cancer state become prominent. These two paths illustrate that the formation of CSCs is roughly from two sources. One route for generating CSCs is from the somatic stem cells with self-renewal capabilities which have the potential to divide into both stem cells and specialized somatic cells, which destined to stop proliferating or die\cite{lobo2007biology}. If these stem cells are out of control for stopping division, but still keeping the ability of self-renewal and differentiation, they become cancer stem cells\cite{ponti2005isolation,ye2015epithelial}. Another route for generating CSCs is that a minor proportion of cancer cells have the capacity of self-renewal and differentiation in their progeny\cite{liu2015epigenetic}. Many experiments have demonstrate that the terminally differentiated cancer cells can gain SCs properties under specific epigenetic conditions\cite{Tang2012Understanding}. These stem cell-like cancer cells drive the cell growth and metastasis are considered as cancer stem cells. Many reports have shown that cancer cells undergoing EMT can obtain stem cell-like characteristics\cite{Mani2008The}, which demonstrate the connection between EMT and CSC. These have been found in hematopoietic\cite{Bonnet1997Human} and solid tumor such as brain\cite{Singh2004Identification} and breast cancer\cite{Alhajj2003Prospective}. These two paths driving the CSCs have the capacity of self-renewal, differentiation and migration independently. The kinetic paths of the landscape view can illustrate the dynamic transitions of SC, CSC and cancer. Due to these diversifications, the CSCs show a series headache ability of drug resistance and recurrence.

From the landscape view, we can also see that there is not only one kinetic path from Normal to Cancer state. There are at least three major paths. One is from SC to CSC to Cancer state. The SCs can gain cancer characteristics and become CSCs. Recently, some studies tracing of CD133+ cells provided direct evidence that SCs were susceptible to cancerous transformation\cite{Zhu2016Multi,Medema2013Cancer}. The CSCs inherit many characteristics of SCs, including self-renewal and differentiation. Moreover, the CSCs with cancerization characteristics such as uncontrollable growth and metastasis. The CSCs can asymmetricly divided into cancer cells and CSCs\cite{Sell2004Stem}. So the CSCs can be seen as the seeds of cancer cells. Another path is from Normal to Premalignant to Cancer state. This can be seen as a cancerous process. On this path, the gene expression level of P53 decreases and ZEB increases. This indicates that the cells not only have pathologic change trend but also have metastasis characteristics at the mean time. The third one is from Normal to Lesion to Hyperplasia to Cancer state. This can be seen as a process which develops from normal cells, going through lesion, gaining the ability of proliferate(hyperplasia) then turning malignant, and eventually falling into Cancer state. Some experiments have show that the lesion often happened before the hyperproliferative changes\cite{Jeremy2003Inflammatory}. The hyperplasia is accumulated to a certain degree, cells process the ability of metastasis, transform to Cancer state ultimately. These three paths would roughly answer a central question in cancer biology, how and which cells can be transformed to cancer. This helps us understanding why cancer is so difficult to cure because the formation of paths.

\subsection{Barrier heights and flux of kinetic paths}

We calculated the barrier heights and the flux of each kinetic paths for further analysis. We know that the barrier heights determine the stability of the states. The higher the barrier is, the more difficult it is to transform from one cell type to another. The barrier heights can also correlate with the switching frequency from one state to another.

\begin{figure}[!ht]
\includegraphics[width=0.50\textwidth]{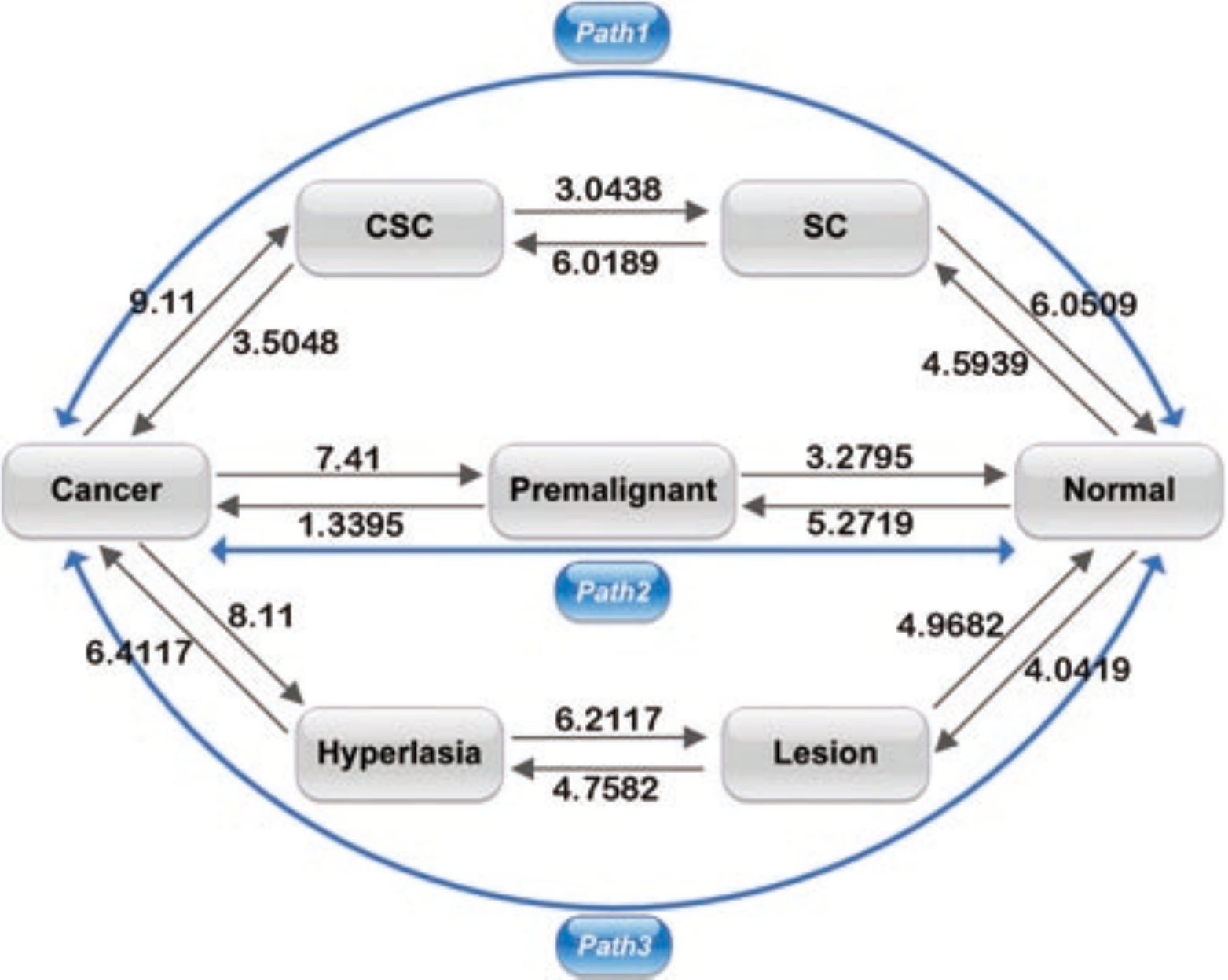}
\caption{The barrier heights between normal, premalignant, cancer, sc, csc, lesion and hyperplasia states and optimal paths among those states. Black arrows represent the barrier from one state to another. The data marked on is the barrier height to overcome. The blue arrows represent the kinetic paths from normal leading to cancer state and the reverse back.} \label{barrier}
\end{figure}

Fig.\ref{barrier} shows that the barrier heights between Normal, Premalignant, Cancer, SC, CSC, Lesion and Hyperplasia states. In $path1$, the barrier height from SC to CSC state is $6.0189$ and from CSC to Cancer state is $3.5048$. We can describe the carcinogenesis of stem cells a high barrier move difficult to occur. It demands strong  regulation and environmental conditions to make the stem cells cancerous\cite{Tang2012Understanding}. The cancer stem cells differentiate to cancer cells with a much lower barrier is a easy process as the CSCs can generate cancer cells progeny when they divided. A cancer stem cell can be asymmetric divided into a cancer cell and a cancer stem cell\cite{Sell2004Stem}. The barrier height from SC to Normal is $6.0509$ which is comparable to the barrier from SC to CSC state. SC state has two choices: to become normal differential cell or become cancer stem cell, both with certain degree of difficulties. In adults, the somatic stem cells are always dormant, it requires specified condition to induce the stem cells to divide. On the other hand, reprogramming requires specific gene regulations. Therefore the barriers for both the differentiation and reprogramming back are relatively high.  When the stem cells are activated the stem cells are asymmetric divided into stem cells and normal somatic cells. It appears that at SC state, the cell can either switch to differentiated cell or cancer stem cell. The paths connect CSC to SC state and CSC to Cancer state with the barriers of $3.0348$ and $3.5048$. It can illustrate that when the cells stay in CSC state, they are both very unstable due to the low barrier height and more likely to transform to Cancer state or back to SC. The barrier height from Cancer to CSC state is $9.11$.  That is also a high barrier which means it is difficult for the cancer cells to transform back to CSCs. Experiments have revealed that there are only a minor proportion of cancer cells having the capacity of self-renewal and differentiation in their progeny\cite{liu2015epigenetic}. Therefore, the switching from cancer cells to CSCs is not very easy to be realized. We can state
that $path1$ has both the characteristics of stem cells and metastasis. Cells going through $path1$ from normal state can change stemness and metastasis, and eventually reach Cancer state.

On $path2$, The barrier heights of Premalignant to Normal state and Normal to Premalignant state are $3.2795$ and $5.2719$ respectively. The barrier height from Premalignant to Normal state is lower than the barrier from Normal to Premalignant. It illustrates that if the cell is in Normal state it is relative difficult to transform to Premalignant state. And if the cell stays in Premalignant state, it is easier to reverse back to Normal state. Moreover, the barrier height of Premalignant state to Cancer and Cancer state to Premalignant are $1.3395$ and $7.41$ respectively. It illustrates that the cell in Premalignant state with intermediate level of metastasis is much easier to transform to Cancer state than reverse from Cancer state back to Premalignant. The barriers between Premalignant state and Normal or Cancer states are lower, so the cell state can transform to the Normal or Cancer state relatively easily. The fatal fate of cancer is its uncontrolled diffusion and metastasis, if the cells are in Cancer state, the metastasis is obvious. So the Premalignant state with intermediate level of tumor and metastasis has vital clinical significance of early diagnosis and prevention of cancer as cells in Premalignant state can transform to Cancer or reverse back to Normal state easily. Therefore, $path2$ has the characteristics of metastasis. Cells going through $path2$ can reflect metastasis process as the Premalignant state is an intermediate state of metastasis. The importance of Premalignant state was discussed in our previous study\cite{yu2016physical}.

In $path3$, we can see the barrier heights between Normal, Lesion and Hyperplasia are $4.0419-4.9682$ and $4.7582-6.2117$. The barriers are not too high. It means that it is not very difficult to transform from one to another. Experiments have shown that the lesion is of common occurrence before the hyperproliferative changes\cite{Jeremy2003Inflammatory}. But the barrier heights between Hyperplasia and Cancer is $6.4117-8.11$. It means that transformation from Hyperplasia to Cancer or reverse back from Cancer to Hyperplasia are both difficult and the Cancer to Hyperplasia state is difficult to occur. That tells us that it is not difficult for the cells transfer from one to another state before the metastasis (transform to Cancer state). If the cells have not reached metastasis, it is relatively easy to cure (reverse to Normal state). And the cells become cancerous due to hyperplasia is a difficult process but it is more difficult to escape from Cancer state to Hyperplasia as it needs to overcome a very high barrier. Therefore, $path3$ reflects the process of accumulated cell damages to metastasis.
Cells in $path3$ going through more and more pathological changes and eventually reach Cancer state.

In Fig.\ref{barrier}, we can see that the paths connect Cancer state to CSC, Premalignant and Hyperplasia state are all with relative high barriers which are $9.11$, $7.41$ and $8.11$ respectively. These illustrate that the barriers of Cancer state are all very high. So the cells in Cancer state are difficult to escape and that can explain why the cancer are so difficult to cure.

\begin{figure}[!ht]
\includegraphics[width=0.40\textwidth]{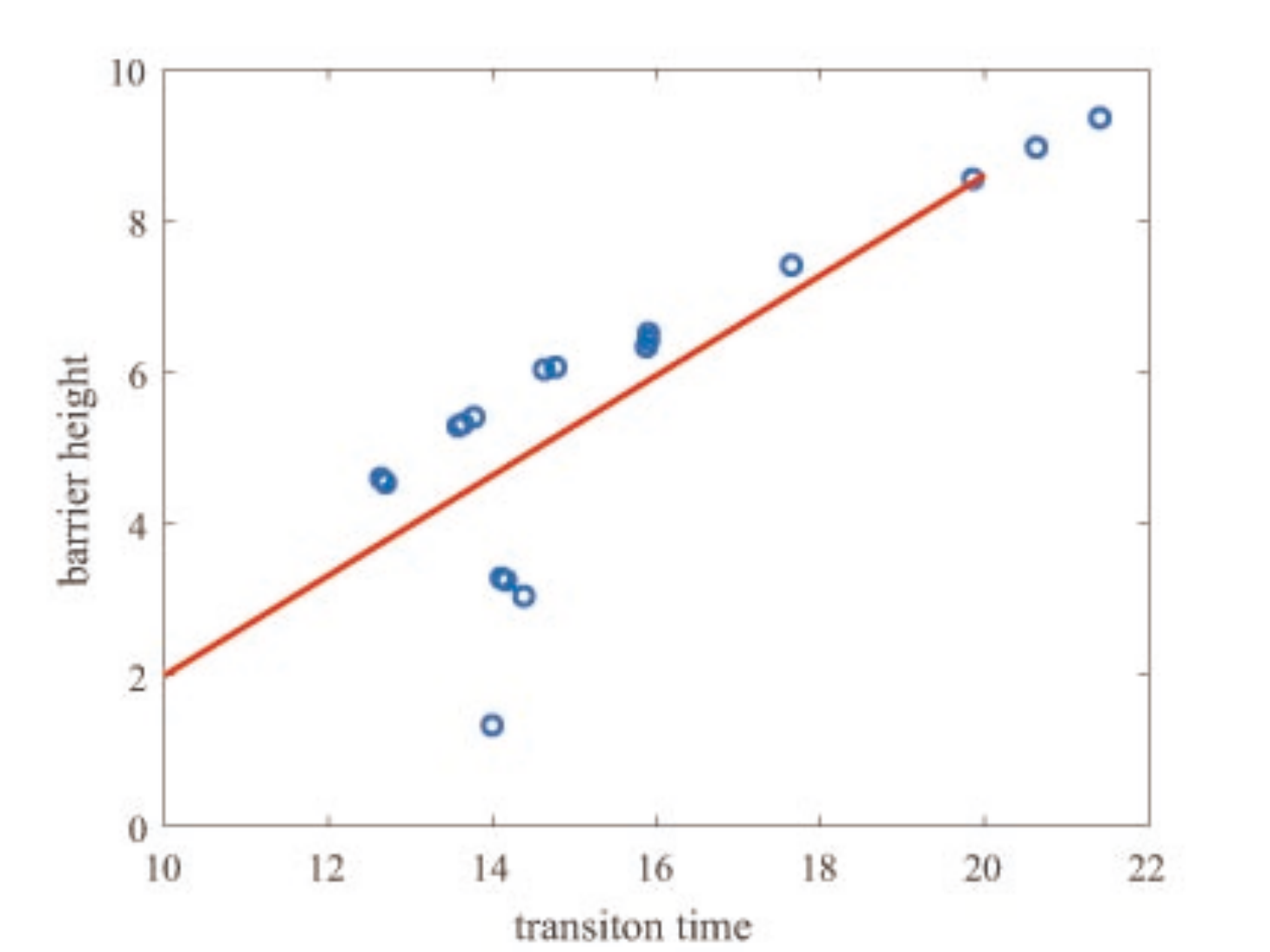}
\caption{The correlation of the transition time and barrier heights. The y-axis represents the data of barrier height and the x-axis represents the data of transition time which have taken $ln$.} \label{corr}
\end{figure}

We have also calculated the correlation of the transition time and the barrier height. The correlation coefficient we calculated is 0.80. As Fig.\ref{corr} shows, the transition time and the barrier height presenting basically the same trend.

We also compared the flux of the three paths (from Normal to Cancer state and the reverse back). The flux of each path can help us to figure out which path is more important in cancer formation. According to the transition time and the probability of each pathway, we can quantify the flux of each path. The transition time in our work depends on the landscape topography. The barrier heights of each state can reflect the landscape topography. The transition rate $k$ of a pathway is the reciprocal of transition time. The details and data can be seen in the support Information. This method has been applied in our previous work\cite{Wang2013-84}. The flux of the path from Normal $\rightarrow$ SC $\rightarrow$ CSC $\rightarrow$ Cancer is $2.2157*10^{-10}$. The probability of this path is $ 0.0719$. The flux of the path from Normal $\rightarrow$ Premalignant $\rightarrow$ Cancer is $2.6227*10^{-9}$. The probability of this path is $0.8509$. The flux of the path from Normal $\rightarrow$ Lesion $\rightarrow$ Hyperplasia $\rightarrow$ Cancer is $2.3813*10^{-10}$. The probability of this path is $0.0773$. The flux and the probability of the path from Normal $\rightarrow$ Premalignant $\rightarrow$ Cancer account for the vast majority of the three. So this path is dominant for Normal to Cancer state transition. Therefore, to prevent cancer formation, this is worthy attention and we have demonstrated the importance of Premalignant state for cancer prevention in our previous work\cite{yu2016physical}.

In the same way, we can also quantify the flux from Cancer to Normal state. The flux of the path from Cancer $\rightarrow$ CSC $\rightarrow$ SC $\rightarrow$ Normal is $2.1830*10^{-9}$. The probability of this path is $0.4243$. The flux of the path from Cancer $\rightarrow$ Premalignant $\rightarrow$ Normal is $9.3693*10^{-10}$. The probability of this path is $0.1821$. The flux of the path from Cancer $\rightarrow$ Hyperplasia $\rightarrow$ Lesion $\rightarrow$ Normal is $2.0245*10^{-9}$. The probability of this path is $0.3935$. The flux of path1( from Cancer $\rightarrow$ CSC $\rightarrow$ SC $\rightarrow$ Normal) and path3 (Cancer $\rightarrow$ Hyperplasia $\rightarrow$ Lesion $\rightarrow$ Normal) are very close which are higher than path2 (Cancer $\rightarrow$ Premalignant $\rightarrow$ Normal). To reverse Cancer state back to Normal, these two paths should be pay attention to. And the flux and probability of Path1 is higher than Path3, so this path is dominant for Cancer to Normal state transition. We need to pay more attention on CSC for cancer curing.

These three paths can address a central question in cancer biology, how and which cells can be transformed to cancer in a quantitative way. The barrier heights can describe the basin depths of the landscape and help us understanding the tendency of the cells transformation from one state to another. Furthermore, the barrier heights of Cancer state are all very high, that means the cells in Cancer state are difficult to transform to others. The presence of multiple cancer formation paths can explain the various mechanisms of cancer formation and that is one of the reasons why cancer is difficult to prevent. The flux of the paths can lead us to find out which path is dominant in the cancer formation and help us describe the difficulty of curing cancer in a quantitative way.

\subsection{Finding key regulations by global sensitivity analysis of landscape topography}
To get further insight of the cancer formation and curing, we explored network motif to find the key regulations by global sensitivity analysis on landscape topography. In the network motif, each gene and regulation contributes to network dynamics. Variation the regulation strengths will influence the barrier heights between attractor basins. Through this way we can figure out which regulations are more sensitive for cancer curing in the network motif.

\begin{figure*}[!ht]
\includegraphics[width=17.4cm,height=8cm]{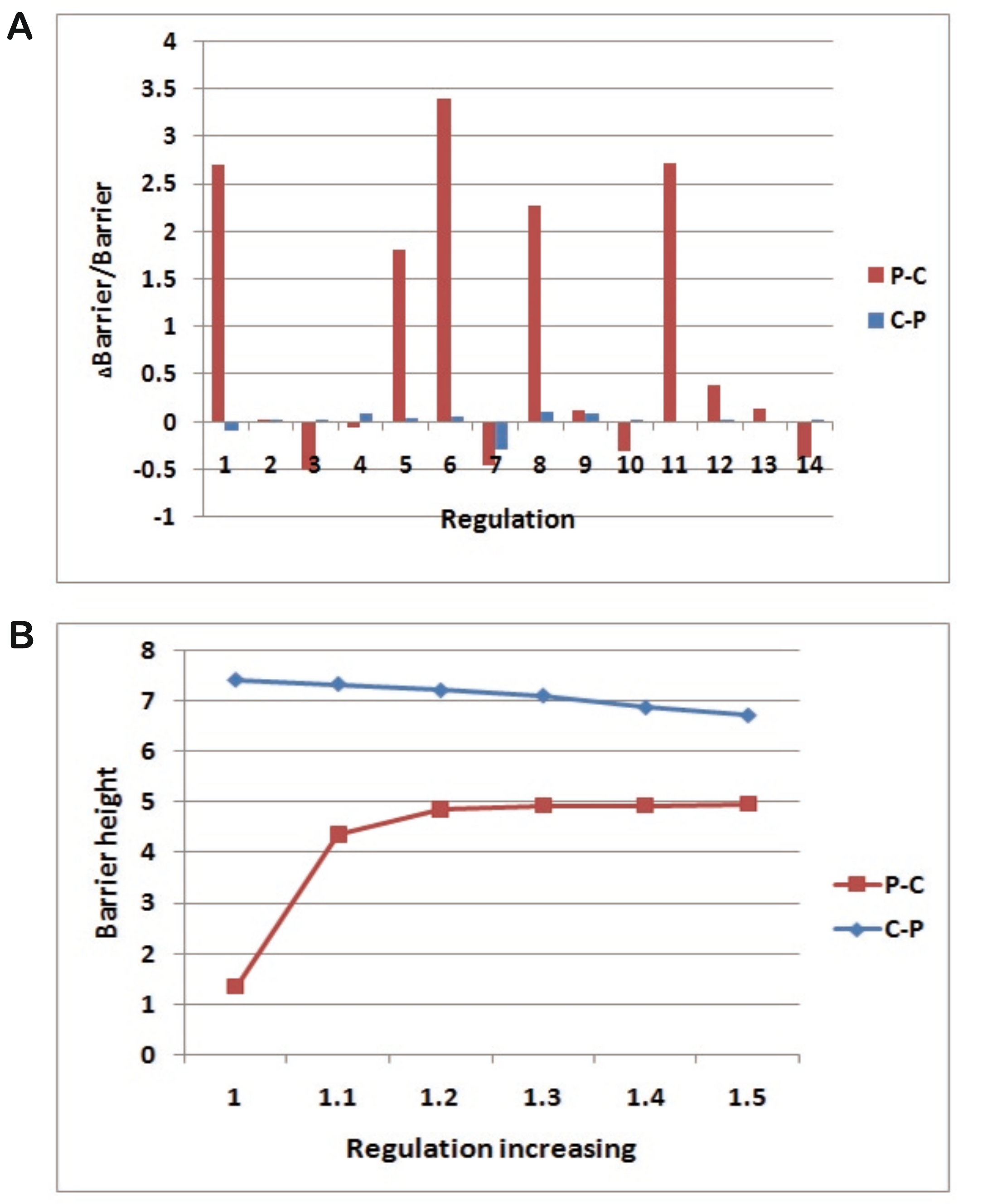}
\caption{Variation rate of barrier height with regulation strength changed. $P-C$ ($C-P$) denotes the barrier height from Premalignant state to Cancer state (Cancer state to Premalignant state). $N-P$ ($P-N$) denotes the barrier height from Normal state to Premalignant state (Premalignant state to Normal state). $N-SC$ ($SC-N$) denotes the barrier height from Normal state to SC state (SC state to Normal state).} \label{barr}
\end{figure*}

Fig.\ref{barr}(A) and (B) display the variation of regulation miR200$\dashv$ ZEB and regulation$1$ is miR200$\dashv$ ZEB in Fig.\ref{barr}(A). Fig.\ref{barr}(C) and (D) display the variation of regulation OCT4$\rightarrow$ OCT4 and regulation$1$ is OCT4$\rightarrow$ OCT4 in Fig.\ref{barr}(C). Fig.\ref{barr}(E) and (F) display the variation of regulation P53$\rightarrow$ P53 and regulation$1$ is P53$\rightarrow$ P53 in Fig.\ref{barr}(E). In Fig.\ref{barr}(A), (C) and (E), the control regulations $2-13$ are P53$\rightarrow$ miR200, P53$\rightarrow$ miR145, P53$\rightarrow$ MDM2, miR145$\dashv$ ZEB, miR145$\dashv$ OCT4, miR145$\dashv$ MDM2, ZEB$\dashv$ miR200, ZEB$\dashv$ miR145, ZEB$\rightarrow$ ZEB, OCT4$\rightarrow$ miR200, OCT4$\rightarrow$ miR145 and MDM2$\dashv$ P53, respectively.

In Fig.\ref{barr}(A) we increased the regulations strength to $1.5$ times. We can see that, in regulation1(miR200$\dashv$ ZEB) the barrier height from Premalignant state to Cancer state increased significantly, and the barrier height from Cancer state to Premalignant state slightly decreased. Although regulation 6 changed very significant too, we abandon it as it changed in the same direction. As we know, the gene ZEB is an EMT activator gene, when the gene expression level is high the metastasis becomes obvious. So when we increased suppression strength of ZEB, the expression level becomes lower than before which leads to the weaker metastasis than before. In that case, the cell state moves from Premalignant to Cancer state much more difficult than before, and moves from Cancer to Premalignant state easier which is beneficial to cancer recovery. In Fig.\ref{barr}(B), we varied the regulation strength from $0.8-1.5$ times. When the regulation strength becomes smaller, the barrier height from Premalignant to Cancer state decreased and the barrier height from Cancer to Premalignant state increased. This could also illustrate the regulation is associated with the variation of the barrier height between Premalignant and Cancer state. This variation of regulation miR200$\dashv$ ZEB can give the information of how to control the metastasis.

In Fig.\ref{barr}(C) we increased the regulations strength to $1.3$ times. OCT4 is a signature
gene of stem cell. If the expression level of OCT4 is high, the stemness of the cell is obvious. We can see that, when the regulation strength increased, the expression level of OCT4 is increased, the barrier height from Normal to SC state is decreased and the barrier height from SC to Normal state increased significantly. This means that the cell in Normal state becomes easier to move to SC state, and the cells in SC state becomes more difficult to move to Normal state. In the meantime, the barrier height from Normal to Premalignant becomes lower than before, and the barrier height from Premalignant to Normal become higher. This indicates that the cell in Normal state is easier to become cancerous, and the cell in Premalignant becomes more difficult to move back. These variation results are consistent with the experiment induced pluripotent stem cells (iPS). Many studies have pointed out that the cellular reprogramming of iPS often lead to the cell with cancerous characteristics which eventually reaches to Cancer state\cite{Kondo2000Oligodendrocyte,Pavlova2016The}. In Fig.\ref{barr}(D), when the regulation strength decreased to $0.7$ times, the barrier height from Normal to Premalignant state is increased and the barrier height from Premalignant to Normal state is decreased. This indicates that the cells in Normal state are more stable, and the cells in Premalignant state are more likely to transform to Normal state. The barrier height from Normal to SC state increased and the barrier height from SC to Normal state decreased. That means the cells in Normal state are harder to switch to SC state and the cells in SC state are more apt to transform to Normal. The regulation OCT4$\rightarrow$ OCT4 can reflect the connection between the stem cells and metastasis. This variation may help researchers finding treatments of cancer through the stem cell clue.

In Fig.\ref{barr}(E), we also increased the regulations strength to $1.3$ times. We can see that the variation of regulation1 (P53$\rightarrow$ P53) can lead the barrier height from Normal to SC state to become higher and the barrier height from SC to Normal state is lower. This indicates that when the expression level of P53 increasing, the cells in Normal state become more difficult to transform to SC state and the cells in SC state is much easier to transform to Normal state. Experiments have shown that P53 is a major driving force for the differentiation of embryonic stem cells (ESCs). Spontaneous differentiation of hESCs reduced significantly when P53 expression is decreased\cite{Qin2007Regulation}. P53 also can provides an effective barrier for the generation of stemness cells from terminally differentiated cells\cite{Solozobova2011p53}. The variation of regulation P53$\rightarrow$ P53 can help us to realize the importance of P53 not only for cancer and but also for stem cell processes.

As we know P53 is a tumor suppressor which we can see in Fig.\ref{barr}(F). When the regulation strength decreased to $0.9$ times and P53 abundance reduced, the barrier height from Normal to Premalignant state has barely changed, but the barrier height from Premalignant to Normal state is increased significantly. In this situation, the cells in Premalignant state are more difficult to move back to Normal state. When the regulation strength is increased to $1.1$ times, the barrier height from Normal to Premalignant state has no significant change and the barrier height from Premalignant to Normal becomes lower. That means the cells in Premalignant state will be easier to Normal state. When the regulation strength is increased to $1.2$ and $1.3$ times, the variation of the barrier height between Normal and Premalignant are both only slightly. When the concentration of P53 reaches a very high level, its characteristic of tumor suppressor become less obvious and other characteristics are present such as inducing the cells to apoptosis\cite{Haupt2003Apoptosis}.

\begin{figure*}[!ht]
\includegraphics[width=16.4cm,height=5cm]{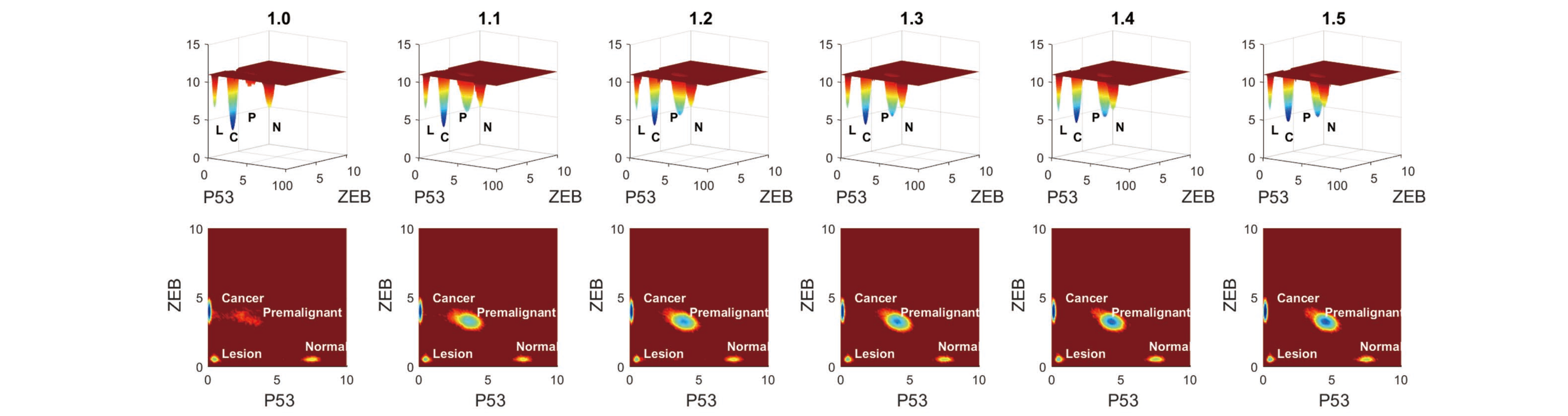}
\caption{Variation of barrier height when the regulation strength changed from $1.0$ to $1.5$ times. Label $L$ represents the Lesion state, Label $P$ represents the Premalignant state, Label $C$ represents the Cancer state and the Label $N$ represents the Normal state.} \label{varr}
\end{figure*}

To see the variation clearly, we depict the landscape topography of miR200$\dashv$ ZEB from regulation strength $1.0$ to $1.5$ times. In Fig.\ref{varr}, the depth of the basin of Premalignant state increased significantly when the regulation strength increased and the depth of the basin of Cancer state is decreased.

\section{Conclusion}

Cancer is a complex and fetal disease. The cancer has features of metastasis, drug resistance and recurrence. These are related to CSCs which leads the cancer to be a major health threat of human being. Recent studies showed that EMT plays a vital role to induce early step of metastasis and is also a way for non-stem cells turning into stem cells. In this study, we develop a dynamic model which includes specific genes and microRNAs for CSC, EMT and cancer, aiming at uncovering the connections among cancer, metastasis and development/differentiation through CSC, EMT and cancer. We quantify the underlying landscape to explore the cancer development/differentiation and metastasis. The origin of cancer can then be elucidated. The kinetic paths and barrier heights between each state can be quantified. The barrier heights determine the stability of the state. Based on the barrier heights, we can relate to the switching frequency of the cells from one state to another. Multiple cancer formation paths can be observed. The flux of each path (from normal leading to cancer and the reverse back) is calculated by the statistics of the path transitions. This is used to figure out which path is more important in cancer formation and cancer curing. This can also help us quantifying the degree of difficulty in curing cancer. Furthermore, we use global sensitivity analysis to find key regulations which are vital for cancer curing. Three regulations miR200$\dashv$ ZEB, OCT4$\rightarrow$ OCT4 and P53$\rightarrow$ P53 are more sensitive to the cancer curing. This work study the functional dynamics and physical mechanisms of cancer development/differentiation and metastasis in a quantitative way. This can give us a guide in cancer clinical therapy.
\section{Acknowledgement}
This study is supported by NSFC grant no.91430217, ,MOST£¬China£¬Grant No.2016YFA0203200 and grant no.NSF-PHY-76066

\bibliographystyle{unsrt}
\bf\scriptsize \linespread{0.5} \setlength{\bibsep}{0.0ex}
\bibliography{cancer_7}

\begin{thebibliography}{10}

\bibitem{gillespie1977exact}
Daniel~T Gillespie et~al.
\newblock Exact stochastic simulation of coupled chemical reactions.
\newblock {\em J. phys. Chem}, 81(25):2340--2361, 1977.

\bibitem{Wang2011-8257}
J.~Wang, K.~Zhang, L.~Xu, and E.~Wang.
\newblock Quantifying the waddington landscape and biological paths for
  development and differentiation.
\newblock {\em Proceedings of the National Academy of Sciences of the United
  States of America}, 108(20):8257--8262, 2011.

\bibitem{Xu2014Exploring}
L.~Xu, K.~Zhang, and J.~Wang.
\newblock Exploring the mechanisms of differentiation, dedifferentiation,
  reprogramming and transdifferentiation.
\newblock {\em Plos One}, 9(8):e105216, 2014.

\bibitem{Sell2004Stem}
S~Sell.
\newblock Stem cell origin of cancer and differentiation therapy.
\newblock {\em Critical Reviews in Oncology/hematology}, 51(1):1--28, 2004.

\bibitem{Takahashi2007-861}
Kazutoshi Takahashi, Koji Tanabe, Mari Ohnuki, Megumi Narita, Tomoko Ichisaka,
  Kiichiro Tomoda, and Shinya Yamanaka.
\newblock Induction of pluripotent stem cells from adult human fibroblasts by
  defined factors.
\newblock {\em Cell}, 131(5):861--872, 2007.
\newblock Times Cited: 7930.

\bibitem{Takahashi2006-663}
Kazutoshi Takahashi and Shinya Yamanaka.
\newblock Induction of pluripotent stem cells from mouse embryonic and adult
  fibroblast cultures by defined factors.
\newblock {\em Cell}, 126(4):663--676, 2006.
\newblock Times Cited: 10050.

\bibitem{Cowin2005Cadherins}
Pamela Cowin, Tracey~M Rowlands, and Sarah~J Hatsell.
\newblock Cadherins and catenins in breast cancer.
\newblock {\em Current Opinion in Cell Biology}, 17(5):499--508, 2005.

\bibitem{Muller2013p53}
Patricia A.~J. Muller and Karen~H. Vousden.
\newblock p53 mutations in cancer.
\newblock {\em Nature Cell Biology}, 15(1):2, 2013.

\bibitem{Martincorena2015Somatic}
I~Martincorena and P.~J. Campbell.
\newblock Somatic mutation in cancer and normal cells.
\newblock {\em Science}, 349(6255):1483, 2015.

\bibitem{Hanahan2000-57}
D.~Hanahan and R.~A. Weinberg.
\newblock The hallmarks of cancer.
\newblock {\em Cell}, 100(1):57--70, 2000.

\bibitem{Lynch1998Mutation}
Michael Lynch, Leigh Latta, Justin Hicks, and Matthew Giorgianni.
\newblock Mutation, selection, and the maintenance of life-history variation in
  a natural population.
\newblock {\em Evolution}, 52(3):727--733, 1998.

\bibitem{Kauffman1971Differentiation}
S~Kauffman.
\newblock Differentiation of malignant to benign cells.
\newblock {\em Journal of Theoretical Biology}, 31(3):429--451, 1971.

\bibitem{Spano2012Molecular}
D~Spano, C~Heck, Antonellis~P De, G~Christofori, and M~Zollo.
\newblock Molecular networks that regulate cancer metastasis.
\newblock {\em Seminars in Cancer Biology}, 22(3):234, 2012.

\bibitem{Blancafort2013Writing}
P~Blancafort, J.~Jin, and S~Frye.
\newblock Writing and rewriting the epigenetic code of cancer cells: from
  engineered proteins to small molecules.
\newblock {\em Molecular Pharmacology}, 83(3):563, 2013.

\bibitem{Rodr2011Cancer}
M~Rodr¨ªguez-Paredes and M~Esteller.
\newblock Cancer epigenetics reaches mainstream oncology.
\newblock {\em Nature Medicine}, 17(3):330, 2011.

\bibitem{Liu2008BRCA1}
Suling Liu, Christophe Ginestier, Emmanuelle Charafe-Jauffret, Hailey Foco,
  Celina~G. Kleer, Sofia~D. Merajver, Gabriela Dontu, and Max~S. Wicha.
\newblock Brca1 regulates human mammary stem/progenitor cell fate.
\newblock {\em Proceedings of the National Academy of Sciences of the United
  States of America}, 105(5):1680--5, 2008.

\bibitem{Tang2012Understanding}
Dean~G Tang.
\newblock Understanding cancer stem cell heterogeneity and plasticity.
\newblock {\em Cell research}, 22(3):457--472, 2012.

\bibitem{hewitt1952transplantation}
HB~Hewitt.
\newblock Transplantation of mouse sarcoma with small numbers of single cells.
\newblock {\em Nature}, 170(4328):622--623, 1952.

\bibitem{Dragu2015Therapies}
Denisa~L Dragu, Laura~G Necula, Coralia Bleotu, Carmen~C Diaconu, and Mihaela
  Chivu-Economescu.
\newblock Therapies targeting cancer stem cells: Current trends and future
  challenges.
\newblock {\em World Journal of Stem Cells}, 7(9):1185, 2015.

\bibitem{Kondo2000Oligodendrocyte}
Toru Kondo and Martin Raff.
\newblock Oligodendrocyte precursor cells reprogrammed to become multipotential
  cns stem cells.
\newblock {\em Science}, 289(5485):1754--7, 2000.

\bibitem{Pavlova2016The}
N.~N. Pavlova and C.~B. Thompson.
\newblock The emerging hallmarks of cancer metabolism.
\newblock {\em Cell Metabolism}, 23(1):27, 2016.

\bibitem{morel2012emt}
Anne-Pierre Morel, George~W Hinkal, Cl{\'e}mence Thomas, Fr{\'e}d{\'e}rique
  Fauvet, St{\'e}phanie Courtois-Cox, Anne Wierinckx, Mojgan
  Devouassoux-Shisheboran, Isabelle Treilleux, Agn{\`e}s Tissier, Baptiste
  Gras, et~al.
\newblock Emt inducers catalyze malignant transformation of mammary epithelial
  cells and drive tumorigenesis towards claudin-low tumors in transgenic mice.
\newblock {\em PLoS Genet}, 8(5):e1002723, 2012.

\bibitem{wang2015links}
Sha-sha Wang, Jian Jiang, Xin-hua Liang, and Ya-ling Tang.
\newblock Links between cancer stem cells and epithelial--mesenchymal
  transition.
\newblock {\em OncoTargets and therapy}, 8:2973, 2015.

\bibitem{scheel2012cancer}
Christina Scheel and Robert~A Weinberg.
\newblock Cancer stem cells and epithelial--mesenchymal transition: concepts
  and molecular links.
\newblock In {\em Seminars in cancer biology}, volume~22, pages 396--403.
  Elsevier, 2012.

\bibitem{mani2008epithelial}
Sendurai~A Mani, Wenjun Guo, Mai-Jing Liao, Elinor~Ng Eaton, Ayyakkannu
  Ayyanan, Alicia~Y Zhou, Mary Brooks, Ferenc Reinhard, Cheng~Cheng Zhang,
  Michail Shipitsin, et~al.
\newblock The epithelial-mesenchymal transition generates cells with properties
  of stem cells.
\newblock {\em Cell}, 133(4):704--715, 2008.

\bibitem{morel2008generation}
Anne-Pierre Morel, Marjory Li{\`e}vre, Cl{\'e}mence Thomas, George Hinkal,
  St{\'e}phane Ansieau, and Alain Puisieux.
\newblock Generation of breast cancer stem cells through epithelial-mesenchymal
  transition.
\newblock {\em PloS one}, 3(8):e2888, 2008.

\bibitem{chaffer2011normal}
Christine~L Chaffer, Ines Brueckmann, Christina Scheel, Alicia~J Kaestli,
  Paul~A Wiggins, Leonardo~O Rodrigues, Mary Brooks, Ferenc Reinhardt, Ying Su,
  Kornelia Polyak, et~al.
\newblock Normal and neoplastic nonstem cells can spontaneously convert to a
  stem-like state.
\newblock {\em Proceedings of the National Academy of Sciences},
  108(19):7950--7955, 2011.

\bibitem{li2015quantifying}
Chunhe Li and Jin Wang.
\newblock Quantifying the landscape for development and cancer from a core
  cancer stem cell circuit.
\newblock {\em Cancer research}, 75(13):2607--2618, 2015.

\bibitem{yu2016physical}
Chong Yu and Jin Wang.
\newblock A physical mechanism and global quantification of breast cancer.
\newblock {\em PloS one}, 11(7):e0157422, 2016.

\bibitem{wellner2009emt}
Ulrich Wellner, J{\"o}rg Schubert, Ulrike~C Burk, Otto Schmalhofer, Feng Zhu,
  Annika Sonntag, Bettina Waldvogel, Corinne Vannier, Douglas Darling, Axel
  Zur~Hausen, et~al.
\newblock The emt-activator zeb1 promotes tumorigenicity by repressing
  stemness-inhibiting micrornas.
\newblock {\em Nature cell biology}, 11(12):1487--1495, 2009.

\bibitem{kumar2012acquired}
Suresh~M Kumar, Shujing Liu, Hezhe Lu, Hongtao Zhang, Paul~J Zhang, Phyllis~A
  Gimotty, Matthew Guerra, Wei Guo, and Xiaowei Xu.
\newblock Acquired cancer stem cell phenotypes through oct4-mediated
  dedifferentiation.
\newblock {\em Oncogene}, 31(47):4898--4911, 2012.

\bibitem{liu2015epigenetic}
Hui-xin Liu, Xiao-li Li, and Chen-fang Dong.
\newblock Epigenetic and metabolic regulation of breast cancer stem cells.
\newblock {\em Journal of Zhejiang University Science B}, 16(1):10--17, 2015.

\bibitem{Yang2013-impact}
P.~Yang, C.~W. Du, M.~Kwan, S.~X. Liang, and G.~J. Zhang.
\newblock The impact of p53 in predicting clinical outcome of breast cancer
  patients with visceral metastasis.
\newblock {\em Scientific Reports}, 3, 2013.

\bibitem{lin2012reciprocal}
Yuanji Lin, Ying Yang, Weihua Li, Qi~Chen, Jie Li, Xiao Pan, Lina Zhou,
  Changwei Liu, Chunsong Chen, Jianqin He, et~al.
\newblock Reciprocal regulation of akt and oct4 promotes the self-renewal and
  survival of embryonal carcinoma cells.
\newblock {\em Molecular cell}, 48(4):627--640, 2012.

\bibitem{lamouille2014molecular}
Samy Lamouille, Jian Xu, and Rik Derynck.
\newblock Molecular mechanisms of epithelial--mesenchymal transition.
\newblock {\em Nature reviews Molecular cell biology}, 15(3):178--196, 2014.

\bibitem{Jeremy2003Inflammatory}
A.~H. Jeremy, D.~B. Holland, S.~G. Roberts, K.~F. Thomson, and W.~J. Cunliffe.
\newblock Inflammatory events are involved in acne lesion initiation.
\newblock {\em Journal of Investigative Dermatology}, 121(1):20, 2003.

\bibitem{rizzino2013concise}
Angie Rizzino.
\newblock Concise review: The sox2-oct4 connection: Critical players in a much
  larger interdependent network integrated at multiple levels.
\newblock {\em Stem Cells}, 31(6):1033--1039, 2013.

\bibitem{lobo2007biology}
Neethan~A Lobo, Yohei Shimono, Dalong Qian, and Michael~F Clarke.
\newblock The biology of cancer stem cells.
\newblock {\em Annu. Rev. Cell Dev. Biol.}, 23:675--699, 2007.

\bibitem{ponti2005isolation}
Dario Ponti, Aurora Costa, Nadia Zaffaroni, Graziella Pratesi, Giovanna
  Petrangolini, Danila Coradini, Silvana Pilotti, Marco~A Pierotti, and
  Maria~Grazia Daidone.
\newblock Isolation and in vitro propagation of tumorigenic breast cancer cells
  with stem/progenitor cell properties.
\newblock {\em Cancer research}, 65(13):5506--5511, 2005.

\bibitem{ye2015epithelial}
Xin Ye and Robert~A Weinberg.
\newblock Epithelial--mesenchymal plasticity: a central regulator of cancer
  progression.
\newblock {\em Trends in cell biology}, 25(11):675--686, 2015.

\bibitem{Mani2008The}
S.~A. Mani, W.~Guo, M.~J. Liao, E.~N. Eaton, A~Ayyanan, A.~Y. Zhou, M~Brooks,
  F~Reinhard, C.~C. Zhang, and M~Shipitsin.
\newblock The epithelial-mesenchymal transition generates cells with properties
  of stem cells.
\newblock {\em Cell}, 133(4):704, 2008.

\bibitem{Bonnet1997Human}
D~Bonnet and J.~E. Dick.
\newblock Human acute myeloid leukemia is organized as a hierarchy that
  originates from a primitive hematopoietic cell.
\newblock {\em Nature Medicine}, 3(7):730--7, 1997.

\bibitem{Singh2004Identification}
Singh SK; Hawkins C; Clarke ID; Squire JA; Bayani J; Hide T; Henkelman RM;
  Cusimano MD;~Dirks PB.
\newblock Identification of human brain tumour initiating cells.
\newblock {\em Nature}, 432(7015):396--401, 2004.

\bibitem{Alhajj2003Prospective}
M~Alhajj, M.~S. Wicha, A~Benitohernandez, S.~J. Morrison, and M.~F. Clarke.
\newblock Prospective identification of tumorigenic breast cancer cells.
\newblock {\em Proceedings of the National Academy of Sciences of the United
  States of America}, 100(7):3983--3988, 2003.

\bibitem{Zhu2016Multi}
Liqin Zhu, David Finkelstein, Culian Gao, Lei Shi, Yongdong Wang, Dolores
  L¨®pez-Terrada, Kasper Wang, Sarah Utley, Stanley Pounds, and Geoffrey Neale.
\newblock Multi-organ mapping of cancer risk.
\newblock {\em Cell}, 166(5):1132, 2016.

\bibitem{Medema2013Cancer}
J.~P. Medema.
\newblock Cancer stem cells: the challenges ahead.
\newblock {\em Nature Cell Biology}, 15(4):338--344, 2013.

\bibitem{Wang2013-84}
Y.~Wang, L.~F. Gan, E.~K. Wang, and J.~Wang.
\newblock Exploring the dynamic functional landscape of adenylate kinase
  modulated by substrates.
\newblock {\em Journal of Chemical Theory and Computation}, 9(1):84--95, 2013.

\bibitem{Qin2007Regulation}
H.~Qin, T.~Yu, T~Qing, Y.~Liu, Y.~Zhao, J.~Cai, J.~Li, Z.~Song, X.~Qu, and
  P.~Zhou.
\newblock Regulation of apoptosis and differentiation by p53 in human embryonic
  stem cells.
\newblock {\em Journal of Biological Chemistry}, 282(8):5842, 2007.

\bibitem{Solozobova2011p53}
Valeriya Solozobova and Christine Blattner.
\newblock p53 in stem cells.
\newblock {\em World Journal of Biological Chemistry}, 2(9):202--214, 2011.

\bibitem{Haupt2003Apoptosis}
Susan Haupt, Michael Berger, Zehavit Goldberg, and Ygal Haupt.
\newblock Apoptosis - the p53 network.
\newblock {\em Journal of Cell Science}, 116(20):4077--4085, 2003.

\end{thebibliography}
\clearpage

\clearpage

\clearpage

\clearpage

\clearpage

\clearpage

\clearpage

\end{document}